\documentclass
[twocolumn,aps,prb,showpacs,superscriptaddress,10pt,citeautoscript]{revtex4-1}%
\usepackage{graphicx}
\usepackage{dcolumn}
\usepackage{bm}
\usepackage{color}
\usepackage{dcolumn}
\usepackage{amsmath}%
\usepackage{amsfonts}%
\usepackage{amssymb}
\usepackage[normalem]{ulem}
\providecommand{\U}[1]{\protect\rule{.1in}{.1in}}

\begin{document}
\title{The role of defects in the metal-insulator transition in VO$_2$ and V$_{2}$O$_{3}$}
\author{Darshana Wickramaratne}
\affiliation{NRC Research Associate, Resident at Center for Computational Materials Science, US Naval Research Laboratory, Washington,
D.C. 20375, USA}
\author{Noam Bernstein}
\affiliation{Center for Computational Materials Science, US Naval Research Laboratory, Washington, DC 20375, USA}
\author{I. I. Mazin}
\affiliation{Center for Computational Materials Science, US Naval Research Laboratory, Washington, DC 20375, USA}
\date{\today}

\begin{abstract}
The vanadates VO$_2$ and V$_2$O$_3$ are prototypical examples of strongly
correlated materials that exhibit a metal-insulator transition.  While the
phase transitions in these materials have been studied extensively, there is a limited
understanding of how the properties of these materials are affected by the 
presence of defects and doping.  In this study we investigate the impact of native
point defects in the form of Frenkel defects
on the structural, magnetic and electronic properties of VO$_2$ and V$_2$O$_3$,
using first-principles calculations. 
In VO$_2$ the vanadium Frenkel pairs lead to a non-trivial insulating state.  The unpaired
vanadium interstitial bonds to a single dimer, which leads to a trimer that has one singlet state
and one localized single-electron $S=1/2$ state.  The unpaired broken dimer
created by the vanadium vacancy also has a localized $S=1/2$ state.  Thus, the insulating
state is created by the singlet dimers, the trimer and the two localized $S=1/2$ states.
Oxygen Frenkel pairs, on the other hand, lead to a metallic state in VO$_2$, but are expected to
be present in much lower concentrations.
In contrast, the Frenkel defects in V$_2$O$_3$ do not directly
suppress the insulating character of the material.  However, the disorder created by defects in
V$_2$O$_3$ alters the local magnetic moments and in turn reduces the energy cost of a transition
between the insulating and conducting phases of the material.  
We also find self-trapped small polarons in  V$_2$O$_3$, which has implications
for transport properties in the insulating phase.
\end{abstract}
\maketitle

\section{Introduction}
\label{sec:intro}
V$_{2}$O$_{3}$ and VO$_{2}$ are prototypical strongly
correlated materials that exhibit a metal-insulator transition (MIT)
\cite{rice1970metal, park2000spin}. The MIT in
V$_{2}$O$_{3}$ and VO$_{2}$ manifests itself as a change in resistivity
of several orders of magnitude. Furthermore, the phase transition in these
materials is accompanied by a concomitant structural (and magnetic in the case
of V$_{2}$O$_{3}$) transition.  At a temperature of 155~K V$_{2}$O$_{3}$ transitions from
a high-temperature (HT) paramagnetic metallic corundum phase 
to a low-temperature (LT) antiferromagnetic insulating monoclinic phase. 
VO$_{2}$, in contrast, is non-magnetic, and exhibits a
transition from a HT metallic rutile structure 
to a LT insulating monoclinic structure at 
340~K\cite{morin1959oxides}.  The interplay between the electronic,
structural and magnetic degrees of freedom in these materials has led to a
number of studies that have sought to exploit the sensitivity of these phase
transitions to the application of strain \cite{wu2006strain}, pressure \cite{mcwhan1973metal}, doping
\cite{mcwhan1971electronic, mcwhan1969mott} and the introduction of defects
\cite{appavoo2012role, Ramirez_2015}.

Despite several decades of research, the ability to manipulate the transition
temperature in these materials remains a challenge due to the lack of a
 quantitative description of how the MIT responds to external stimuli. 
This is reflected in the results of recent experiments\cite{Ramirez_2015}, 
where the response of the MIT to ion irradiation was shown to
be qualitatively different in V$_2$O$_3$ as compared with VO$_2$.
V$_{2}$O$_{3}$ and VO$_{2}$ thin films were subjected to ion irradiation at
different dosage levels. The electronic properties of the two 
materials showed markedly different properties pre and post irradiation. 
Prior to irradiation, both
vanadate thin films exhibited a MIT as evidenced by a large
change in electrical resistivity across the
respective transition temperature of the two materials.
Post-irradiation, the transition temperature was observed to decrease in
V$_{2}$O$_{3}$ while the magnitude of the resistivity on both sides of the 
transition temperature remained unchanged
compared to the unirradiated sample. At a critical irradiation dosage, the MIT in V$_{2}$O$_{3}$ was
completely suppressed and the electrical resistivity exhibited metallic-like
conduction. In contrast, the MIT phenomenon in VO$_{2}$ appeared to be more
robust. The transition temperature post-irradiation remained unchanged compared
to the unirradiated sample while the resistivity below the transition
temperature decreased by approximately two orders of magnitude. At higher
dosages of irradiation, evidence of a MIT in VO$_{2}$ remained, although
the magnitude of the resistivity change between the insulating and conducting
state was lower compared to the unirradiated sample. In both vanadates, there
was no evidence of structural distortion or strain introduced as a
result of the irradiation. 
Hence, it is likely that any changes
observed in the MIT phenomenon in the two materials is due to the introduction of
defects. A crucial question then is, what is the microscopic origin of the
different response of the MIT in both materials to the presence of defects?

Point defects can affect the electronic structure through a multitude of mechanisms. One
possibility is that the defects act as sources of free carriers which would
lead to appreciable conductivity below the MIT temperature and a collapse in
the MIT. Another possibility is that the presence of defects introduces structural disorder. A
\textit{quantitative} description of the impact of these effects on the MIT
in the vanadates will need to accurately account for the role of
strong correlations and the structural, electronic and magnetic properties of
each material.

The above overview makes it clear that aspects related to the role of
defects in the vanadates remain to explored. Motivated by this, we use
first-principles density functional theory (DFT) calculations to determine the electronic, magnetic and
structural properties of defects in V$_{2}$O$_{3}$ and VO$_{2}$. 

\subsection{General considerations}
\label{sec:general}
The chemical simplicity of V$_{2}$O$_{3}$ and VO$_{2}$ led to 
a majority of the initial research effort devoted to these materials
to assume an oversimplified view of their electronic structure, 
where V$_{2}$O$_{3}$ was treated as a prototypical Mott insulator
and VO$_{2}$ as a Peierls insulator.
However, intensive and decades-long research has shattered
this deceptive simplicity and an understanding has gradually emerged that (i)
the MIT in the two vanadates have very different properties and (ii) in each case,
several different physical phenomena contribute. In the following we will
present this understanding and how it has emerged on a qualitative level (as a
result of numerous quantitative calculations with complementary methodologies).

\begin{figure}[h]
\includegraphics[width=8.5cm]{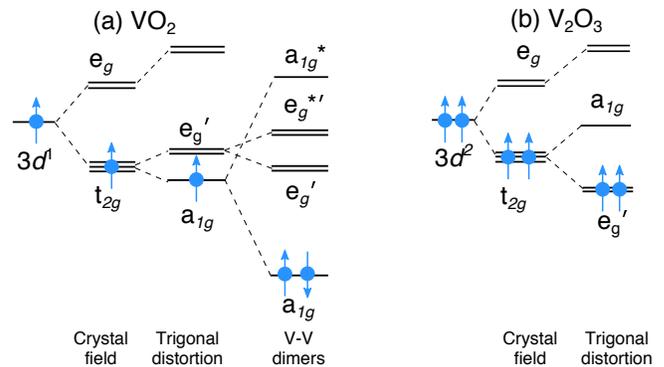}\caption{Schematic of the energy levels of vanadium
 $d$-states and their occupation in
 (a) VO$_2$ and (b) V$_2$O$_3$.  For VO$_2$ the single $3d^{1}$ state subject to an octahedral
crystal field, trigonal distortion and the energy separation between the bonding
$a_{1g}$ and antibonding $e_{g}^{\pi}$ orbitals due to covalent bonds between V dimers 
is illustrated.  For V$_{2}$O$_{3}$ the ordering of the two $3d^{2}$ states subject to an octahedral crystal
field and trigonal distortion is illustrated.}
\label{fig:cf}%
\end{figure}

In VO$_{2},$ the vanadium ion is in a 4$^{+}$ oxidation state, that is, it has 1 $d$-electron.
The HT crystal structure of VO$_{2}$ has a high
symmetry rutile structure, which does not allow for an insulator, except for a
Mott insulator. Oxygen forms nearly perfect octahedra around the V ions,
 thus splitting the V $d$ states into a well separated $t_{2g}$ triplet and 
$e_{g}$ doublet (cf. Fig.~\ref{fig:cf}(a)).
First-principles calculations show that a Hubbard parameter $U-J\gtrsim$ 2~eV opens an
antiferromagnetic Mott-Hubbard gap at $T=0$ \cite{eyert2002metal}.  However, this phase
exists only above 340 K, which is above its intrinsic N\'eel temperature, so experimentally 
it is always paramagnetic and metallic.
Deviation from antiferromagnetic order 
strongly suppresses the tendency toward insulating behavior; indeed, in a ferromagnetic
arrangement the Hubbard gap only opens at a larger value $U-J\gtrsim$ 3.5~eV. It is worth
noting that the electronic structure of VO$_2$ in the metallic phase does not exhibit any nesting
features that would trigger a conventional, Fermi-surface driven
charge density wave and dimerization of the structure. Indeed, the LT structure is not really
reminiscent of a charge density wave, but rather exhibits strongly coupled
dimers with the V-V distance along the $c$-axis changing from uniform separation of 
2.85 \AA~to alternating dimers with bond lengths of 2.65 \AA~and 3.12 \AA, which is also
accompanied by a strong trigonal distortion of approximately 30\% \cite{longo1970refinement}.  
This trigonal distortion splits the nearly-degenerate $t_{2g}$
orbitals \cite{biermann2005dynamical} into an elongated $a_{1g}$ singlet and a more isotropic
$e_{g}^{\prime}$ doublet (cf. Fig.~\ref{fig:cf}(a)). The $a_{1g}$ orbitals in each dimer point toward each
other and form a strong covalent bond \cite{goodenough1960direct}. In an ionic picture, this bond is
occupied by two electrons forming a typical covalent singlet (analogous to a
H$_{2}$ molecule) and a gap opens between this covalent state and the lowest
$e_{g}^{\prime}$ state (cf. Fig.~\ref{fig:cf}(a))

In a solid, these states broaden into bands.  Describing these bands within DFT
leads to bandwidths that are overestimated and their relative separation is underestimated, a 
manifestation of the well known 
``band gap'' problem in DFT.
While there are many remedies to this effect, such as DFT+$U$, dynamical mean field theory (DMFT),
hybrid functionals or the recently developed SCAN functional, which indeed all open a gap in VO$_2$
\cite{eyert2002metal, kylanpaa2017accuracy}, all of them simultaneously increase the propensity
toward magnetism and disrupt the formation of singlet dimers (DFT-like theory can only
describe a single-determinant state, and not a true singlet). {\em Cluster} dynamical mean field theory (DMFT), 
on the contrary, has this capability, but is too computationally expensive to be applied to large 
supercells\cite{biermann2005dynamical,weber2012vanadium}. 
Fortunately, non-magnetic GGA calculations, as discussed in more detail
below, yield a structure rather close to the experimentally observed structure,
only slightly overestimating the degree of dimerization.  This behavior can be used, in some cases, to 
circumvent the problem by optimizing the atomic structure using one method and interrogating the
electronic structure using another.

The case of V$_{2}$O$_{3}$ is more subtle. The difference is that V is in the $3^{+}$ oxidation state, which
has two $d$-electrons
that would not fit into one covalent bond.  Thus, the dimerization mechanism
described above is not operative (note that Ti$_2$O$_3$, having one electron less per Ti atom, {\it does} dimerize
at low temperatures\cite{chang2018c}).
Prior studies have demonstrated
\cite{park2000spin, rodolakis2010inequivalent}
 that the relatively small trigonal distortion 
in this compound also splits $a_{1g}$ and $e_{g}^{\prime}$ states (cf. Fig.~\ref{fig:cf}(b)),
but this separation is considerably smaller than the
bandwidths, at least on a one-electron level.  Hence, in DFT calculations 
it is a metal. 
Importantly, the doubly degenerate $e_{g}^{\prime}$ orbitals are lower in energy
than the $a_{1g}$ ones, so any interaction that amplifies the $e_{g}^{\prime}-a_{1g}$
splitting opens a gap in V$_2$O$_3$ with its 2 $d$-electrons per V.  Since all of the
 relevant orbitals are localized on a single V site, the DFT+$U$ method is well suited 
for describing this effect. Note that the physics that leads to the insulating state here is
very different from that in VO$_{2}$. Furthermore, it was recently shown that the
MIT in V$_{2}$O$_{3}$ in itself is not a Mott transition at all, even though
it is a transition between a Mott insulator and a correlated metal 
\cite{paolasini1999orbital,leiner2018frustrated}.
Rather, it is a strongly-first order transition between two systems that are different
in more aspects than just their electronic structure. Specifically, the
magnetic interactions in the HT corundum phase are strongly frustrated,
and if the structural transition upon cooling could have been arrested, the
materials would have remained paramagnetic (and likely metallic) down to very
low temperatures. Conversely, if the LT monoclinic structure could be
stabilized well above 155 K, it would have remained an ordered Mott
antiferromagnet well above room temperature.  This suggests
that the MIT in V$_{2}$O$_{3}$ should be more sensitive to
structural aspects than in VO$_{2}.$

Some first-principles studies on strongly
correlated oxides have highlighted the
important effect structural disorder/defects may have on the energy barriers 
between different magnetic states \cite{han2018lattice, kim2017magnetic}. 
Similarly, a recent study demonstrated that structural distortions associated
with replacing V with Cr are responsible for enhanced correlation effects and
stabilizing an insulating state at high temperatures.\cite{lechermann2018uncovering}  It was also shown that
despite Cr and V having different number of electrons such a replacement does
not result in charge doping.  Conversely, doping with Ti does result in
charge doping, albeit, counterintuitively, in electron rather than hole doping.
The response of V$_{2}$O$_{3}$ to this charge doping is that expected for a
true Mott insulator, that is, it rapidly transforms into a correlated metal.

\subsection{Outline}

In this study we investigate the points defects that are likely to be
introduced by ion irradiation\cite{Ramirez_2015} by considering the 
properties of a neutral Frenkel defect, i.e. a self-interstitial and a corresponding 
(but not adjacent) vacancy that form due to an atomic displacement, in V$_{2}$O$_{3}$ and VO$_{2}$.
We also examine the role of pure electron and hole doping by introducing
electrons and holes with overall charge neutrality ensured by a uniform
compensating background. Investigating these effects separately allows us to
capture the physics of charge doping in isolation from the effects of atomic
disorder as introduced by the presence of defects.

The relatively complete understanding of the physics of pristine and doped
V$_{2}$O$_{3}$ and VO$_{2}$ suggests a straightforward qualitative
explanation of the qualitatively different responses of the two compounds 
when subject to irradiation.  Irradiation displaces a small fraction of ions.
If we consider the displacement of V ions, this will result in the creation
of vanadium Frenkel pairs, where the diplaced V ions and the newly
formed vacancies are not on nearest-neighbor sites.

In the case of VO$_2$, breaking a V dimer 
results in two unpaired V ions, while,
presumably, leaving the other dimers relatively intact (in the following
sections we will quantify this statement) and they will remain covalent
singlets. The unpaired V ions, assuming that they do not change their valence
(again, we will quantify this statement later), will have one $d$-electron each,
subject to Hubbard correlations, and will form isolated $S=1/2$ impurities,
localized by the Hubbard interaction. The material will thus remain
insulating (but, if it will be possible to measure the magnetic response
sufficiently accurately, we expect the material to exhibit a
 Curie-like spin susceptibilty with an
effective moment of $2\sqrt{S(S+1)}=\sqrt{3}$ $\mu_{B}$ per 
unpaired V atom).  Our calculations confirm these general considerations
 with one additional interesting observation:  the unpaired vanadium
interstitial bonds to one of the dimers forming a covalently
bonded trimer.  This trimer has one localized $S=1/2$ unpaired state
that is subject to Hubbard correlations.  Furthermore, applying a reasonable
$U$ to VO$_2$ with the vanadium Frenkel pair leads to a band gap that is significantly
suppressed compared to the band gap of pristine VO$_2$.  This is consistent with the
experimental observation that upon irradiation of VO$_2$, the MIT 
is preserved but the resistivity in the insulating phase decreases by up to two orders 
of magnitude compared to pristine VO$_2$.
In contrast, we find that oxygen Frenkel pairs lead to a metallic state.  However, the
formation energy of the oxygen Frenkel pairs is high.  Hence, we expect them
to exist at lower concentrations and in turn not affect the macroscopic transport
properties of VO$_2$ significantly.
For the case of VO$_2$ we only focus on the properties of Frenkel defects
in the insulating monoclinic phase since our goal is to address
how robust the insulating state and the vanadium dimerization is to the presence
of native defects.

Our considerations of the electronic structure of V$_2$O$_3$ would suggest
the crystal field, and therefore the valence state, of V in
V$_{2}$O$_{3}$ is rather sensitive to local structural distortions. If (and we
will see that this is indeed the case) displaced V ions in V$_{2}$O$_{3}$
acquire a different valence state, one may expect this to result in charge doping and
presumably a rapid suppression of the MIT temperature,
 analogous to Ti-doped V$_2$O$_3$ \cite{lechermann2018uncovering}. 
Our calculations, presented below, partially confirm this picture, but also
uncover unexpected insight. Most importantly, the calculations suggest
that for a low concentration of Frenkel pairs (as low as in
experiment \cite{Ramirez_2015}), the extra charge does not become itinerant.  Instead, the extra
charge localizes on a single vanadium atom, alters the V-O bond length and
leads to the formation of a localized small polaron.
Furthermore, while the presence of Frenkel defects does not lead to a doping driven suppression of the 
Mott insulating state in V$_{2}$O$_{3}$, we find the Frenkel pairs reduce the energy cost of the transition
between the LT and HT
magnetic states. This reduction in the energy to transition between
the magnetic states is consistent with the experimental observation that in V$_{2}$O$_{3}$ the
MIT temperature decreases upon irradiation.

In Sec.~\ref{sec:methods} we describe the computational methodology. In
Sec.~\ref{sec:electronic} we discuss the structural and electronic properties
of VO$_2$ and V$_{2}$O$_{3}$.
The properties of defects in VO$_2$ are described in Sec.~\ref{sec:vo2} and the
properties of defects in V$_{2}$O$_{3}$ are described in Sec.~\ref{sec:v2O3}.
Key results from our study are summarized
in Sec.~\ref{sec:conclusions}.

\section{Computational Methods}

\label{sec:methods} 
Our calculations are based on density functional theory
within the projector-augmented wave method \cite{Blochl_PAW} as implemented in
the \textrm{VASP} code \cite{VASP_ref,VASP_ref2} using the generalized
gradient approximation defined by the Perdew-Burke-Ernzerhof (PBE) functional
\cite{perdew1996generalized}. In our calculations, \textrm{V} $4s^{2}%
3p^{6}3d^{3}$ electrons and \textrm{O} $2s^{2}2p^{4}$ electrons are treated as
valence. All calculations use a plane-wave energy cutoff of 600~eV. For the
calculations in the primitive unit cells we use
an $8\times8\times8$ $k$-point grid. 
In order to
simulate the Mott-insulating behavior of V$_{2}$O$_{3}$ and to capture the
role of strong correlations in VO$_{2}$ we use a spherically-averaged Hubbard
correction within the fully-localized limit double-counting subtraction
\cite{anisimov1993density}. 
In the case of VO$_2$, we optimize the volume and atomic coordinates of the 
unit cell and the defect supercell using non-spin polarized PBE calculations.  Subsequent
calculations of the VO$_2$ electronic structure that use the PBE optimized structure are
based on DFT+$U$.  For the case of V$_{2}$O$_{3}$, all of our calculations are based on
DFT+$U$.  We apply a $U-J$ value of 1.8~eV
to the V $d$-states for the calculations of V$_{2}$O$_{3}$ and a
$U-J$ value of 2~eV to the V $d$-states in the case of VO$_2$.
 These parameters yield the closest agreement with the experimental band gaps in both materials.

Calculations of defects in both materials relied on 
the supercell approach. For the lowest-energy defect geometries we use  
quasi-cubic supercells with 324 atoms for VO$_{2}$ and 360 atoms
for V$_{2}$O$_{3}$.
In order to scan several less favorable geometries we used 
smaller supercells of 96 atoms for VO$_{2}$ and 160 atoms for V$_{2}$O$_{3}$.
Calculations with these small VO$_2$ and V$_{2}$O$_{3}$
defect supercells were performed on $2\times2\times2$ $k$-point grids
while for the largest supercells only the $\Gamma$-point
was used to optimize the structure and obtain total energies and densities of states. 

We identify the most favorable defect configuration by evaluating several configurations
of Frenkel pairs in each vanadate. To generate the initial geometry
for these configurations we find maximally large spherical voids in the
primitive cell.  For each spherical void we place an interstitial in
the center, and pick the farthest site of the same species (considering
the effects of the periodic boundaries) for the vacancy. We then relax
the configuration with respect to atomic positions and unit cell size and shape.
The resulting formation energy of, for example, an oxygen
Frenkel pair in V$_{2}$O$_{3}$, is defined as:
\begin{equation}
\begin{aligned} E^{f}({\rm O}_{i}{\rm -}v_{\rm O}) = E_{\rm tot}(O_{i}{\rm -}v_{\rm O}) 
- E_{\rm tot}({\rm V_2O_3}) \end{aligned}\label{eq:form}%
\end{equation}
where E$_{\mathrm{tot}}$(O$_{i}$-$v_{\mathrm{O}}$) denotes the total energy of
the V$_2$O$_3$ defect supercell with an oxygen interstitial, O$_{i}$ and an 
oxgen vacancy, $v_{\rm O}$.  E$_{\mathrm{tot}}%
$(\textrm{V$_{2}$O$_{3}$}) is the total energy of the pristine V$_{2}$O$_{3}$
supercell. Since we only consider Frenkel pairs in this study, the formation
energy is not dependent on the atomic chemical potential. 
We also consider the formation energy of charged Frenkel pairs in the singly negative
and positive charge states and find them to be higher in energy than the neutral
Frenkel pair. 
Unless otherwise
stated, we only report on the results of geometries for neutral Frenkel pairs in each material 
that have the lowest formation energy.

We also investigate the formation of self-trapped small polarons in pristine V$_2$O$_3$. 
In the case of small hole polarons we remove an electron from the
pristine supercell of each material, perturb the initial magnetic moment on a single
V atom, decrease the V-O bond length around the single V atom with respect to the
equilibrium V-O bond length and then allow for a complete relaxation of all atomic coordinates. 
 In the case of small
electron polarons we add an electron to the defect supercell, perturb the 
magnetic moment around a single V atom and increase the V-O bond length around the V site
before allowing for a complete relaxation of all atomic coordinates.
Note that investigation of self-trapped polarons in VO$_2$ is not possible with the DFT+$U$
calculations used in this study, for the technical 
reasons outlined in Sec.~\ref{sec:general}.  Namely, small polarons must be magnetic and 
require full account of Hubbard correlations, but any account of correlations at the 
level below cluster DMFT overestimates the tendency to magnetism and kills the dimerization.

\section{Results}

\subsection{Defect-free structures}
\label{sec:electronic} 
\begin{figure}[h]
\includegraphics[width=8.5cm]{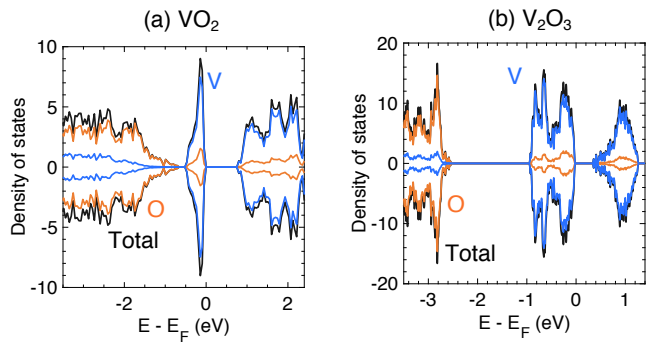}\caption{Density of states for
the monoclinic insulating phase of (a) VO$_2$ and (b) V$_2$O$_3$.  The majority
spin states are illustrated with positive values and the minority spin states are illustrated
with negative values}
\label{fig:dos}%
\end{figure}

We first report on the electronic structure of VO$_2$.
For the LT monoclinic phase of VO$_2$, we find 
that a non-spin-polarized PBE-functional relaxation leads
to good agreement with experiment for both the unit cell lattice constants and V dimerization distance.
The relaxed lattice constants are $a$ = 5.631 \AA, $b$ = 4.541 \AA~and
$c$ = 5.254 \AA, in close agreement with the experimentally
reported lattice constants for monoclinic VO$_2$:
$a$ = 5.750 \AA, $b$ = 4.540 \AA~and~$c$ = 5.380 \AA \cite{longo1970refinement}.
The V atoms form dimers where the V short bond length is 2.515~{\AA}, 
consistent with x-ray diffraction measurements
of 2.62~{\AA}\cite{longo1970refinement}.  In contrast, relaxing the atomic
coordinates and the volume of the VO$_2$ monoclinic unit cell
using DFT+$U$ enhances the tendency for the
V atoms to become magnetic and leads to equally spaced V atoms along
the $c$-axis.  However, the electronic structure from a DFT+$U$ calculation 
of a monoclinic VO$_2$ structure optimized using a non-spin polarized PBE calculation yields
good agreement with experiment.
The $U-J$ value of 2~eV we
use in our calculations leads to a band gap of 0.68~eV, which is in
close agreement with the experimental band gap of 0.60~eV determined by
infra-red absorption \cite{qazilbash2008electrodynamics} measurements.
The density of states (DOS) of VO$_{2}$ in its insulating monoclinic phase is
plotted in Fig.~\ref{fig:dos}(a).  The valence band is comprised of 
hybridized V $d$-states and O $p$-states while the bottom of the conduction band 
primarily consists of V $d$-states.

For V$_2$O$_3$ we investigate the high-temperature (HT) and low-temperature (LT) phases.
To describe the
antiferromagnetic configuration of the LT monoclinic phase
of V$_{2}$O$_{3}$ we use a four formula-unit cell.  We find the ground state
magnetic order to be one where the magnetic moments are ferromagnetically
aligned within the $a$-$c$ plane of the monoclinic structure and are antiferromagnetically
aligned along the monoclinic $b$ axis, consistent with neutron scattering studies
of V$_2$O$_3$ and previous calculations in the insulating phase \cite{leiner2018frustrated, moon1970antiferromagnetism}. 
The atomic coordinates, shape and volume of the unit cell are optimized using DFT+$U$
with the parameters outlined in Sec.~\ref{sec:methods}. We find the lattice
constants of the LT monoclinic structure of V$_{2}$O$_{3}$ to be $a$ = 7.414
\AA , $b$ = 5.084 \AA ~and~$c$ = 5.559 \AA , which is in agreement with the experimental LT
lattice constants ($a$ = 7.255 \AA , $b$ = 5.002 \AA ~and $c$ = 5.548 \AA )
reported for monoclinic V$_{2}$O$_{3}$ \cite{dernier1970crystal}. The density
of states for the LT monoclinic phase of V$_{2}$O$_{3}$ is plotted in
Fig.~\ref{fig:dos}(b).  The top of the valence band in V$_{2}$O$_{3}$ 
is comprised of V $d$-states with a minor contribution from O $p$-states.
The bottom of the conduction band is composed entirely of V $d$-states.
Our DFT+$U$ calculations of monoclinic V$_2$O$_3$ 
yield a band gap of 0.35~eV, which is close to the band gap of $\sim$0.40~eV
obtained from optical conductivity measurements in V$_2$O$_3$ \cite{qazilbash2008electrodynamics}.

We use a two formula-unit
cell for the HT metallic corundum phase of V$_{2}$O$_{3}$.
Neutron scattering studies have shown this phase to be a highly
frustrated paramagnet \cite{leiner2018frustrated}. Description of such a state is 
always a challenge in DFT, because the existence of the fluctuating local moments
is essential for structural properties (cf. Fe-based superconductors), but 
the standard DFT can only describe either non-magnetic (not {\it para}magnetic!) or
magnetically ordered states. Fortunately, as shown in Ref.~\onlinecite{leiner2018frustrated},
the corundum phase of V$_{2}$O$_{3}$ is magnetically frustrated so magnetic ordering must affect
the total energy only weakly. Of the possible ordered phases we have selected the ferromagnetic one as a
proxy for a magnetically disordered state because it respects the same full lattice symmetry (as opposed to 
antiferromagnetic arrangements).
We use the same $U-J$ value of 1.8 eV as before. 
Optimizing the atomic coordinates, shape and volume of the ferromagnetic corundum unit cell using DFT+$U$
leads to the following lattice constants: $a$ = 5.0375~{\AA} and $c$ = 14.305~{\AA}.  This
is in agreement with the experimental lattice constants,
$a$ = 4.952 \AA~and~$c$ = 14.003 \AA, of the corundum phase of 
V$_2$O$_3$ measured using x-ray diffraction at room temperature \cite{dernier1970crystal}.

For the two different magnetic configurations of the corundum structure
we calculate a spin-flip energy $\Delta E = E_{tot}(AFM) - E_{tot}(FM)$, where $E_{tot}(AFM)$ and $E_{tot}(FM)$ 
are the total energies of the V$_2$O$_3$ supercell with the V atoms antiferromagnetically and 
ferromagnetically aligned, respectively.  Hence, negative spin-flip energies correspond to the antiferromagnetic
configuration being favored.  The spin-flip energy between antiferromagnetic and ferromagnetic phase of
corundum V$_{2}$O$_{3}$ (allowing for full
relaxation of the atomic coordinates, cell shape and volume) is only 0.0037~eV per 
V$_{2}$O$_{3}$ formula unit (or 10 K per V).
This is consistent with the notion that the
system is magnetically frustrated and its ordering temperature would be well 
below the temperature range where it exists in nature, just as 
found by Leiner \textit{et al}.~\cite{leiner2018frustrated}

We note that there is a principal physical difference between the MIT in the two
vanadates. In V$_2$O$_3$ it is controlled by the delicate energy balance between the two
structurally {\it and magnetically} different phases.  Hence, we
focus on the impact defects have on their energies. In 
VO$_2$, the transition occurs at a higher temperature and does not involve magnetic order.
A suppression of insulating behavior there, if any, can only occur 
through metallization of the LT monoclinic phase, so in this study we only concentrated on studying 
this phase in VO$_2$ in the presence of defects.

\subsection{VO$_{2}$ defects}
\label{sec:vo2}

\begin{figure}[h]
\includegraphics[width=8.5cm]{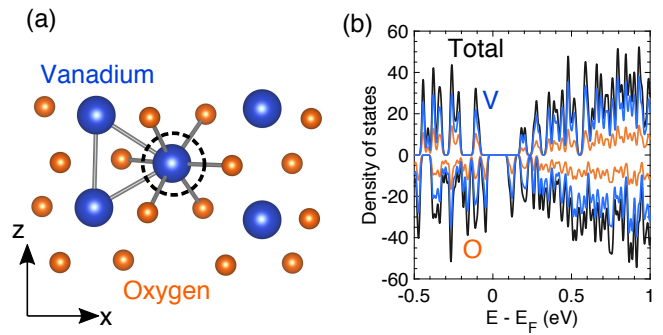}\caption{
(a) Schematic of the VO$_2$ supercell illustrating the vanadium self-interstitial (black dotted circle)
that bonds with one of the V-V dimers.
(b) Projected density of states for the LT monoclinic structure of VO$_2$ with
a V Frenkel pair.
}
\label{fig:vo2-vfrenkel}%
\end{figure}

We consider the properties of V and O Frenkel pairs in
VO$_{2}$, optimizing the defect supercell using non-spin polarized PBE calculations. 
We start with the V Frenkel pair, where we find that the unpaired vanadium interstitial bonds to one of the V-V dimers
forming a nearly equilateral triangle with a V-V bond length of $\sim$2.50 \AA.
Furthermore, as illustrated in Fig.~\ref{fig:vo2-vfrenkel}(a),
the unpaired V interstitial is 
also octahedrally coordinated by the nearest neighbor oxygen atoms; the V-O bond lengths range from
1.89~{\AA} to 2.05~{\AA}, which are close to the V-O bond lengths of pristine VO$_2$.
We investigated several positions of the unpaired vanadium self interstitial and find this
configuration, where it bonds to a single dimer, to be the most stable, with a formation
energy of 3.67 eV.  The V vacancy breaks a dimer leaving behind an unpaired
V atom, and leads to oxygen dangling bonds that disrupt the V dimerization
near the vacancy site.  We therefore report results for large supercells with more than 100 VO$_2$ 
formula units, which maintain the dimerization of the vanadium atoms far from the vacancy site,
unless otherwise noted.

The geometry of the unpaired vanadium interstitial 
can be understood from the following analysis, starting from
a dimer and an isolated ion. As discussed above, the dimer forms a doubly
occupied bonding state at $-t_{dd\sigma}$, where $t_{dd\sigma}$ is the hopping (direct overlap) between $a_{1g}$ orbitals,
and several nonbonding and antibonding states.  Since the electron of the isolated atom is in a nonbonding
state, at this level of approximation the noninteracting electron energy of the three ions in question is $-2t_{dd\sigma}$.
Arranging the ions in a triangle creates one bonding state at $-2t_{dd\sigma}$, in addition to some non-bonding
and anti-bonding states. Populating the bonding state with two electrons and a non-bonding state with
the third electron leads to an electron energy gain of $-4t_{dd\sigma}$.  This configuration is clearly lower
in energy than the original ``dimer+isolated'' arrangement.
Indeed, our DOS calculations show two states with opposite spins, localized on
this triangle, well below the Fermi level. The third electron may either stay on the triangle, in
which case it becomes subject to Hubbard repulsion just as well as the electron localized on the isolated
V ion, and does not contribute to conductivity. It can also move into the $e_g'$ conductivity band and become 
metallic. Which case is realized is impossible to say on the model level, however, our DFT+$U$ calculations,
described below, indicate that a relatively small $U$ readily creates an insulator. 
Indeed, our spin-polarized DFT+$U$ calculations of the electronic structure
of the V Frenkel lead to a gap as seen in Fig. ~\ref{fig:vo2-vfrenkel}(b),
albeit suppressed significantly in comparison to the band gap
of pristine VO$_2$ (cf. Fig.~\ref{fig:dos}(a)).

Thus, we conclude
that upon irradiation the following, rather non-trivial insulating state is created.
In addition to the singlet dimers away from the defects, it consists of one state with $S=1/2$ localized
on the unpaired broken-dimer neighbor of the vacancy, a singlet localized on the V triangle that
includes the interstitial atom, and another $S=1/2$ state localized on the same triangle.
The reduction in the band gap due to the vanadium Frenkel pairs in VO$_2$ would lead to a lower
electrical resistivity in the insulating phase compared to pristine VO$_2$.  We note that this is consistent
with electrical transport measurements on irradiated VO$_2$ \cite{Ramirez_2015}, 
where the resistivity of VO$_2$ in the insulating
phase decreased by up to two orders of magnitude upon irradiation compared to pristine VO$_2$.
 
\begin{figure}[h]
\includegraphics[width=8.5cm]{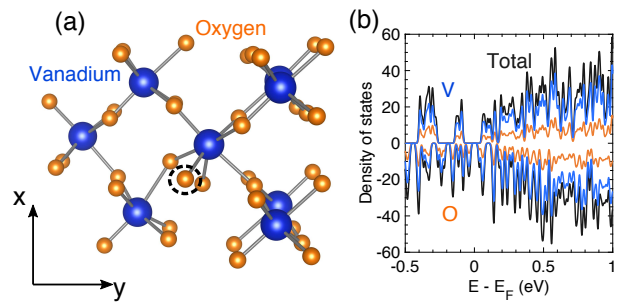}\caption{
(a) Schematic of the VO$_2$ supercell illustrating the oxygen self-interstitial (black dotted circle)
that is a part of the oxygen Frenkel pair defect.
(b) Projected density of states for the LT monoclinic structure of VO$_2$ with
a O Frenkel pair}
\label{fig:vo2-ofrenkel}%
\end{figure}

We also consider the role of oxygen Frenkel pairs in VO$_2$, where
we find that the unpaired oxygen interstitial forms a bond with one vanadium
atom.  This oxygen is located 1.42~{\AA} from one of the other oxygens coordinated
to the same vanadium, forming a dimer (cf. Fig~\ref{fig:vo2-ofrenkel}(a)),
with oxygen-vanadium distances of 1.95~{\AA} and 2.01~{\AA}.
The five remaining oxygens form V-O bonds with bond lengths that range from 1.84~{\AA} to
2.01~{\AA}. This configuration of the oxygen Frenkel leads to a formation
energy of 4.86~eV, lowest among the various oxygen Frenkel configurations that we considered, but 
higher than that of the above-described V Frenkel of 3.67 eV.

O defects disrupt the crystal field of the nearest V ions, which in our calculations
leads to closing of the gap, Fig.~\ref{fig:vo2-ofrenkel}(b). However, the O Frenkel pairs
have a higher formation energy and will have a much lower concentration compared to the 
V Frenkel pairs.  Furthermore, prior studies of ionic transport in semiconducting
transition metal oxides have found the migration barrier for oxygen vacancies to be lower than
the migration barrier of cation vacancies or cation self interstititials
\cite{medasani2017vacancies,medasani2018first}.  Hence, we also expect any O Frenkel pairs
that may form in VO$_2$ to heal faster compared to the V Frenkel pairs.  As a result we
speculate O Frenkel defects will not affect the macroscopic transport properties of VO$_2$.

Thus far we have only considered the properties of the Frenkel defects with the 
lowest formation energy for either type of defects.
 It is conceivable that for the highest irradiation dosages Frenkel defects with higher formation
energies can also be introduced\cite{Ramirez_2015}.  Such high energy defects may have 
electronic properties that are 
different from the lowest energy Frenkel defect configurations that we have considered
thus far. To this end, we also examined the electronic structure of the Frenkel defect configurations
that have higher formation energies than the lowest energy configuration we have considered thus far
(using, as described in Sec.~\ref{sec:methods}, smaller 96 atom supercells).  
In the case of the V Frenkel defects, we 
considered five other configurations of the V vacancy and self-interstitial pair 
that have formation energies up to 1.1~eV higher
than the lowest energy configuration.  We find the same qualitative features
in the electronic and structural properties.  The band gap is reduced with respect to
the pristine material due to each defect.  Due to the small sized supercell, we find the dimerization
of the V atoms is no longer present.  We also examined the electronic structure of five different
configurations of O Frenkel defects that have formation energies up to 1~eV higher
than the lowest energy defect.  For each of these configurations, we also find the 
effects qualitatively similar to those with the lowest formation energies.

\subsection{V$_{2}$O$_{3}$ defects}
\label{sec:v2O3} 
\begin{figure}[h]
\includegraphics[width=8.5cm]{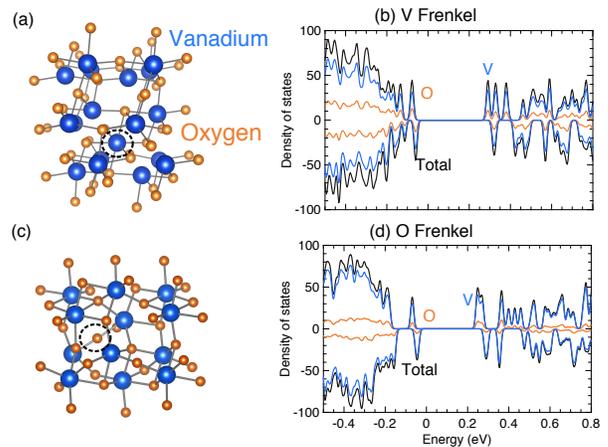}
\caption{
(a) Schematic of the V$_2$O$_3$ supercell illustrating the vanadium self-interstitial (black dotted circle)
that is a part of the vanadium Frenkel pair defect.
(b) Projected density of states for the 
LT antiferromagnetic structure of 
V$_2$O$_3$ with the V Frenkel relative to the Fermi energy.  
(c) Schematic of the V$_2$O$_3$ supercell illustrating the oxygen self-interstitial (black dotted circle)
that is a part of the oxygen Frenkel pair defect.
(d)  Projected density of states for the LT antiferromagnetic structure of
V$_2$O$_3$ with the O Frenkel relative to the Fermi energy.}
\label{fig:v2o3-frenkel}%
\end{figure}

For V$_{2}$O$_{3}$ in the antiferromagnetic or
ferromagnetic configuration the V interstitial 
relaxes to a position between two collinear $c$-axis aligned dimers.
Figure \ref{fig:v2o3-frenkel}(a) schematically illustrates the position of the V
interstitial for the V Frenkel. The V atoms in this pentamer are all
ferromagnetically aligned. The formation energy of the V Frenkel in the
antiferromagnetic V$_{2}$O$_{3}$ supercell is 3.77~eV. We considered the
possibility of the V Frenkel where the spin of the V interstitial is
anti-aligned with the nearest neighbor V atoms and found it to have a 
higher formation energy of 3.98~eV. 
For the V Frenkel in the antiferromagnetic V$_{2}%
$O$_{3}$ supercell, we find the gap to be preserved (cf. Fig.~\ref{fig:v2o3-frenkel}(b)).

We also consider an oxygen Frenkel pair in V$_{2}$O$_{3}$. The oxygen interstitial that is a part of the \textrm{O} Frenkel
pair leads to occupied states that are below the valence band and
unoccupied states that are above the conduction band. The formation energy
of the O Frenkel defect is 6.44~eV, which is higher than the formation energy
of the V Frenkel pair. For the O Frenkel pair in the antiferromagnetic V$_{2}%
$O$_{3}$ supercell, we also find the gap to be preserved and V$_{2}$O$_{3}$
remains insulating (cf. Fig.~\ref{fig:v2o3-frenkel}(d)).

In addition to a possible direct effect on the band gap, defects in V$_2$O$_3$ 
may change the magnetic ordering energies and therefore affect the MIT, since it
is coupled to the structural and magnetic transitions.
For both Frenkel defects, we optimize the atomic coordinates of the V$_2$O$_3$ supercell
in a configuration where the undisplaced V atoms of the V$_2$O$_3$ supercell
are ferromagnetically and antiferromagnetically
aligned (taking into account the relaxed volumes of the two respective magnetic configurations).  
For pristine V$_2$O$_3$, the spin-flip energy, $\Delta E$, is -0.0065~eV per V atom.

The spin-flip energy is -0.0021~eV per V atom in the supercell with a V Frenkel,
and -0.0038~eV per V atom with an O Frenkel.  Thus, the presence of either type
of Frenkel pair defect in V$_2$O$_3$ lowers the spin flip energy relative to the pristine
material.  The results are summarized in
Table~\ref{tab:v2O3_energies}.

\begin{table}[tb]
\caption{Spin-flip energy, $\Delta$E per V atom, between V$_{2}%
$O$_{3}$ in an antiferromagnetic configuration and in a ferromagnetic
configuration obtained with 360 atom supercells.}
\label{tab:v2O3_energies}
\centering
\begin{tabular}
[c]{c|c}\hline\hline
Structure & $\Delta$E per V atom (eV)\\[1.25 ex]\hline
Pristine & -0.0065\\[1.25 ex]%
V Frenkel & -0.0021\\[1.25 ex]%
O Frenkel & -0.0038\\[1.25 ex]\hline\hline
\end{tabular}
\end{table}

It is evident from Fig.~\ref{fig:v2o3-frenkel}(b) and Fig.~\ref{fig:v2o3-frenkel}(d)
that the vanadium and oxygen Frenkel pairs do not suppress the V$_{2}$O$_{3}$ gap.
However, both defects decrease the spin-flip
energy and lower the energy cost of a transition to a metallic
ferromagnetic state. This reduction in the spin-flip energy arises from the
broken connectivity between the V atoms that are aligned antiferromagnetically
along the $b$-axis. The local distortion introduced by the vacancy is
sufficient to modify the magnetic moments of the dangling bond atoms.  Furthermore,
the self-interstitial in both Frenkel defects disrupts the magnetic moment
of the nearest neighbor atoms that it bonds to.  This combination of effects is 
sufficient to lower the spin-flip energy due
to both types of Frenkel pairs that we consider here. 
As we discuss in Sec.~\ref{sec:general} and demonstrate with 
calculations of the spin-flip energy in Sec.~\ref{sec:electronic}, the HT phase of V$_2$O$_3$
is a highly frustrated paramagnet \cite{leiner2018frustrated} with a low N\'eel temperature
while the LT phase is strongly antiferromagnetic with a high N\'eel temperature.
Hence, we suggest that this suppression in the energy cost due to the
presence of defects arises primarily from an energy gain in the LT insulating
phase of V$_2$O$_3$ that contains the Frenkel defects which shifts the energy balance towards the HT phase
and is a plausible source for the reduction of the MIT temperature
upon irradiation of V$_{2}$O$_{3}$ \cite{Ramirez_2015}.

\subsubsection{V$_{2}$O$_{3}$ polarons}
\label{sec:v2o3_polarons}

The disruptions of the bonding caused by the the V and O Frenkel pair
defects in V$_{2}$O$_{3}$ also leads to hole doping due to changes
in the valence state of the defect or near-defect atoms.  However,
these holes do not occupy band states, but instead localize through the formation
of hole polarons. In the case of the V Frenkel pair, we find the formation of
two hole polaron sites centered on two different V
atoms. The effective oxidation state on these V atoms changes from V$^{3+}$ to
V$^{4+}$. The nearest neighbor V-O bond lengths of the hole polaron sites are
3$\%$ shorter than the equilibrium V-O bond length. This raises the question
whether small hole and electron polarons can exist as isolated species in V$_{2}$O$_{3}$. 

\begin{figure}[tb]
\includegraphics[width=8.5cm]{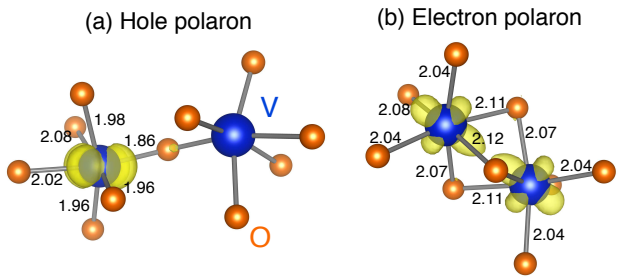}\caption{Atomic configuration
and the charge density isosurface for a small (a) hole polaron and (b)
electron polaron in a V$_{2}$O$_{3}$ supercell.  The V-O bond lengths around the
polaron site are shown alongside each bond in units of {\AA}ngstroms.}%
\label{fig:polaron}%
\end{figure}

To investigate small polarons in V$_2$O$_3$ we apply the approach
described in Sec.~\ref{sec:methods}.  We find small hole polarons to be stable
in V$_2$O$_3$.  The hole polaron localizes on
a single V atom (along with a minor contribution from a nearest neighbor O atom), as
illustrated in Fig.~\ref{fig:polaron}(a). The valence of the V atom changes
from V$^{3+}$ to V$^{4+}$. The formation of a small hole polaron leads to a
distortion of the local geometry: we find the nearest neighbor V-O bonds that
surround the polaron are 3 to 5$\%$ shorter than
the equilibrium V-O bond length. The self-trapping energy for the small hole
polaron (the energy difference between the configuration with the localized
hole and the atomic configuration with a delocalized hole) is 0.12~eV.

We find electrons in the conduction band of V$_{2}$O$_{3}$ can also localize
to form small electron polarons. The localized electron polarons are more
stable than the delocalized electron by 0.065~eV. 
We find the electron localizes on two nearest-neighbor V
atoms that are ferromagnetically aligned, as illustrated in
Fig.~\ref{fig:polaron}(b). This polaron configuration, where the extra electron
resides on V dimers that are ferromagnetically coupled instead of a single V
atom, is often referred to as a Zener polaron \cite{zhou2000zener}. In this
configuration the ferromagnetically coupled V-dimers on which the electron
polaron resides have a bond length that is 3.4$\%$ shorter than the
equilibrium V-V bond length within the $a-c$ plane. We also considered the
possibility of stabilizing a Zener polaron where the two V atoms are aligned
antiferromagnetically. We find this configuration of the self-trapped electron
polaron to be 0.016~eV per V atom higher in energy.

Since $U$, a calculation parameter whose value is only known empirically, 
can affect the energy of the polaron, 
it is pertinent to question if the self-trapped polaron is indeed the ground
state configuration in the case of electron and hole doped V$_{2}$O$_{3}$. To
verify that this is indeed the case we evaluated the self-trapping energy of
small electron and hole polarons in V$_{2}$O$_{3}$ as a function of the
on-site potential $U$ starting from the lowest value of $U$=1.3~eV which leads
to a gap in the V$_{2}$O$_{3}$ monoclinic structure. For values of $U$
from 1.3~eV to 5~eV, we indeed do find small electron and hole polarons to be
more stable compared to the delocalized state.  In the case of the electron polaron
we find a transition from the electron being localized on a V dimer to
being localized on a single V atom. This is accompanied by an
increase in the V-V nearest-neighbor bond length. This is consistent with the
competition between intersite hopping between the ferromagnetically aligned V
dimers and Hund's exchange \cite{streltsov2016covalent} and is an effect that
has been predicted to occur in other transition metal oxides. 

Since we find self-trapped small polarons to be more stable than delocalized electrons or holes,
carrier transport measurements in the insulating state of
V$_{2}$O$_{3}$ would lead to hopping-like transport of
the polarons between neighboring V sites \cite{holstein1959studies}. This
hopping transport of small polarons will have a characteristic activation
energy that can be extracted from the temperature-dependent measurements of
the conductivity.  First-principles calculations of polaron migration barriers
yield an upper limit to the activation energy for transport by small polarons \cite{maxisch2006ab}.
To understand how electron polarons move through the V$_2$O$_3$
lattice, we identify two adjacent pairs of V sites where an electron polaron is stable
and is localized on a V dimer.  We then use a linear interpolation
of the two structures and calculate the total energy of each intermediate structure.
We find the migration barrier for electron polarons to be 0.09~eV.
We apply a similar approach to determine the migration barrier for hole polarons and find
it to be 0.11~eV.  Hence, we suggest temperature-dependent measurements
of carrier transport in the insulating phase of V$_2$O$_3$ would observe evidence of 
hopping transport with an activation energy of $\sim$0.1~eV that can attributed to the
presence of small polarons in the material.

\section{Summary and Conclusions}
\label{sec:conclusions}
In conclusion, we examined the role of Frenkel defects and their influence on the MIT in the
vanadates, in particular, focusing on the observation by Ramirez {\it et al.} \cite{Ramirez_2015} that the MIT collapses
in V$_2$O$_3$ while it remains robust in VO$_2$.

From our calculations, the following scenario emerges:
The recently discovered\cite{leiner2018frustrated} unique property of the MIT in
V$_2$O$_3$ is that the transition occurs between a highly frustrated 
paramagnetic phase with a very low intrinsic N\'eel temperature, and a strongly
antiferromagnetic phase with a very high intrinsic N\'eel temperature, and is therefore
strongly first order. As a result, the transition temperature and its very existence
are highly sensitive to the magnetic ordering energy in the low-temperature phase. We have shown that 
the structural disruptions introduced by defects in the form of Frenkel pairs strongly reduce this energy and thus shift
the energy balance toward the HT metallic phase.
The physics of VO$_2$ is very different. We have shown that the vacancy and interstitial formed when displacing vanadium 
atoms lead to a single undimerized V and a dimer-interstitial trimer,
leaving the dimers away from either defect
intact.  The unpaired electrons that are part of V Frenkel defect are localized, and lead to a small gap when subject
to Hubbard correlations.  We suggest the presence of V Frenkel defects in the insulating phase of 
VO$_2$ would lead to a reduction in the resistivity compared to pristine VO$_2$.
O displacement, according to our calculations closes the band gap of VO$_2$.  However, we find oxygen Frenkel pairs to have
a higher formation energy, at least for the concentrations we considered.  We speculate that
O defects form in a smaller amount (due to a higher formation energy) and also heal faster, so their 
concentration is much smaller than 1\%, as considered in our calculations. The resulting
concentration must be too small to affect macroscopic transport properties of VO$_2$.

Our second result is that in the insulating phase of V$_2$O$_3$ we find small polarons to
be stable, either assisted by the presence of defects or as a self-trapped species.
Self-trapped small polarons in V$_2$O$_3$ can lead to hopping-like conductivity; we find
a migration barrier of 0.09~eV for small hole polarons and 0.11~eV for electron polarons.
We propose temperature-dependent electrical or optical 
conductivity measurements in the insulating phase of V$_2$O$_3$
would be instrumental in elucidating the presence of small polarons in this material.

\acknowledgements
D.W acknowledges support from the National Research Council fellowship at the US Naval Research Laboratory.
The work of N.B and I.I.M was supported by the Laboratory-University Collaboration Initiative of the DoD Basic Research Office.


%

\end{document}